\documentclass[technote,10pt,twocolumn]{IEEEtran}

\usepackage{cite}

\ifCLASSINFOpdf

\else

\fi

\usepackage{amsmath}

\usepackage{amssymb}
\usepackage[bb=dsserif]{mathalpha}

\usepackage{array}
\usepackage{float}
\usepackage{kotex}
\usepackage{graphicx}
\usepackage{bm}
\usepackage{bbm}
\usepackage{verbatim}
\usepackage{svg}

\usepackage{caption,subcaption}

\newsavebox{\measurebox}

\usepackage{algorithm}
\usepackage{algorithmic}
\begin{document}

%
% paper title
% Titles are generally capitalized except for words such as a, an, and, as,
% at, but, by, for, in, nor, of, on, or, the, to and up, which are usually
% not capitalized unless they are the first or last word of the title.
% Linebreaks \\ can be used within to get better formatting as desired.
% Do not put math or special symbols in the title.
\title{Rate-splitting Multiple Access for Hierarchical HAP-LAP Networks under Limited Fronthaul

}
%
%
% author names and IEEE memberships
% note positions of commas and nonbreaking spaces ( ~ ) LaTeX will not break
% a structure at a ~ so this keeps an author's name from being broken across
% two lines.
% use \thanks{} to gain access to the first footnote area
% a separate \thanks must be used for each paragraph as LaTeX2e's \thanks
% was not built to handle multiple paragraphs
%

\author{Jeongbin Kim, Seongah Jeong, \textit{Member, IEEE}, 

Seonghoon Yoo, \textit{Student Member, IEEE}, Woong Son, and Joonhyuk Kang, \textit{Member, IEEE}}

% make the title area
\maketitle

% As a general rule, do not put math, special symbols or citations
% in the abstract or keywords.

\vspace{-0.3cm}
\begin{abstract}
In this correspondence, we propose hierarchical high-altitude platform (HAP)-low-altitude platform (LAP) networks with the aim of maximizing the sum-rate of ground user equipments (UEs). The multiple aerial radio units (RUs) mounted on HAPs and LAPs are managed by the central unit (CU) via constrained fronthaul links. The limitation of fronthaul capacity can be addressed through quantization, employing the cloud radio access network (C-RAN) architecture. For spectral efficiency, we adopt the rate-splitting multiple access (RSMA), leveraging the advantages of both space-division multiple access (SDMA) and non-orthogonal multiple access (NOMA). To achieve this, we jointly optimize rate splitting, transmit power allocation, quantization noise variance, and UAV placement using an alternating optimization (AO) approach coupled with successive convex approximation (SCA) and the weighted minimum mean square error (WMMSE) method. Numerical results validate the superior performance of the proposed method compared to benchmark schemes, including partial optimizations or those without the assistance of LAPs.
\end{abstract}

% Note that keywords are not normally used for peerreview papers.

\IEEEpeerreviewmaketitle

\vspace{-0.1cm}
\section{Introduction}
Unmanned aerial vehicles (UAVs) have garnered significant attention due to their cost-effectiveness and on-demand deployment capabilities. Specifically, UAVs are widely adopted as aerial base stations (UAV-BSs) to improve both throughput and coverage in cellular networks, especially in areas with high traffic congestion. Nevertheless, the performance and endurance of UAV-assisted communication systems are inevitably constrained by the finite onboard energy capacity of UAVs \cite{zeng2016wireless}. In response to this limitation, hierarchical aerial systems with multiple high-altitude platforms (HAPs) and low-altitude platforms (LAPs) have been introduced \cite{kang2023cooperative}. By capitalizing on the wide coverage of HAPs and the high accessibility of LAPs to ground user equipment (UEs), the HAP-LAP networks can meet the quality-of-service (QoS) requirements of users within a predetermined lifetime.

For the improvement of spectral efficiency in UAV communication networks, rate-splitting multiple access (RSMA) \cite{mao2022rate} has been actively studied. RSMA dynamically splits rates based on interference levels, capitalizing on the advantages of both space-division multiple access (SDMA) and non-orthogonal multiple access (NOMA). Recently, the cloud radio access network (C-RAN) architecture \cite{hossain2019recent} has been explored to implement the concept of rate-splitting, which is well-suited for multi-UAV communications. This architecture effectively manages interference, facilitating flexible deployment and expansion of UAVs through centralized processing.

In \cite{jaafar2020downlink}, downlink communications with a single UAV were investigated to maximize the weighted sum-rate by jointly optimizing RSMA parameters and UAV placement. Another proposal \cite{ahmad2019uav} introduces UAV-assisted C-RAN with rate-splitting for base station breakdown scenarios, jointly designing beamformers and rate-splitting to maximize the weighted sum-rate.

Motivated by these works, we propose HAP-LAP networks in which multiple ground UEs communicate with a single HAP, and multiple LAPs function as UAV-BSs using the RSMA technique. In this setup, the HAP and LAPs are managed by the central unit (CU) through constrained fronthaul links, and the limited fronthaul capacity is addressed through quantization employing the C-RAN architecture. Our objective is to maximize the sum-rate of ground UEs, and to achieve this, we jointly optimize rate-splitting, power allocation, quantization noise, and UAV placement. This optimization is based on an alternating optimization (AO) technique coupled with successive convex approximation (SCA)  \cite{boyd2004convex} and the weighted minimum mean square error (WMMSE) \cite{yu2019efficient}.
Through numerical results, we validate the superior performance of our proposed design compared to conventional schemes, such as partial optimizations or single UAV communications. Notably, our work is distinct as it considers rate-splitting for HAP-LAP networks with constrained fronthaul—a perspective that provides valuable insights into practical UAV communications, an aspect not explored in existing literature.

\vspace{-0.1cm}
\section{System Model}

\begin{figure}[t]
  \centering
   \includegraphics[scale=1,width=1\linewidth]{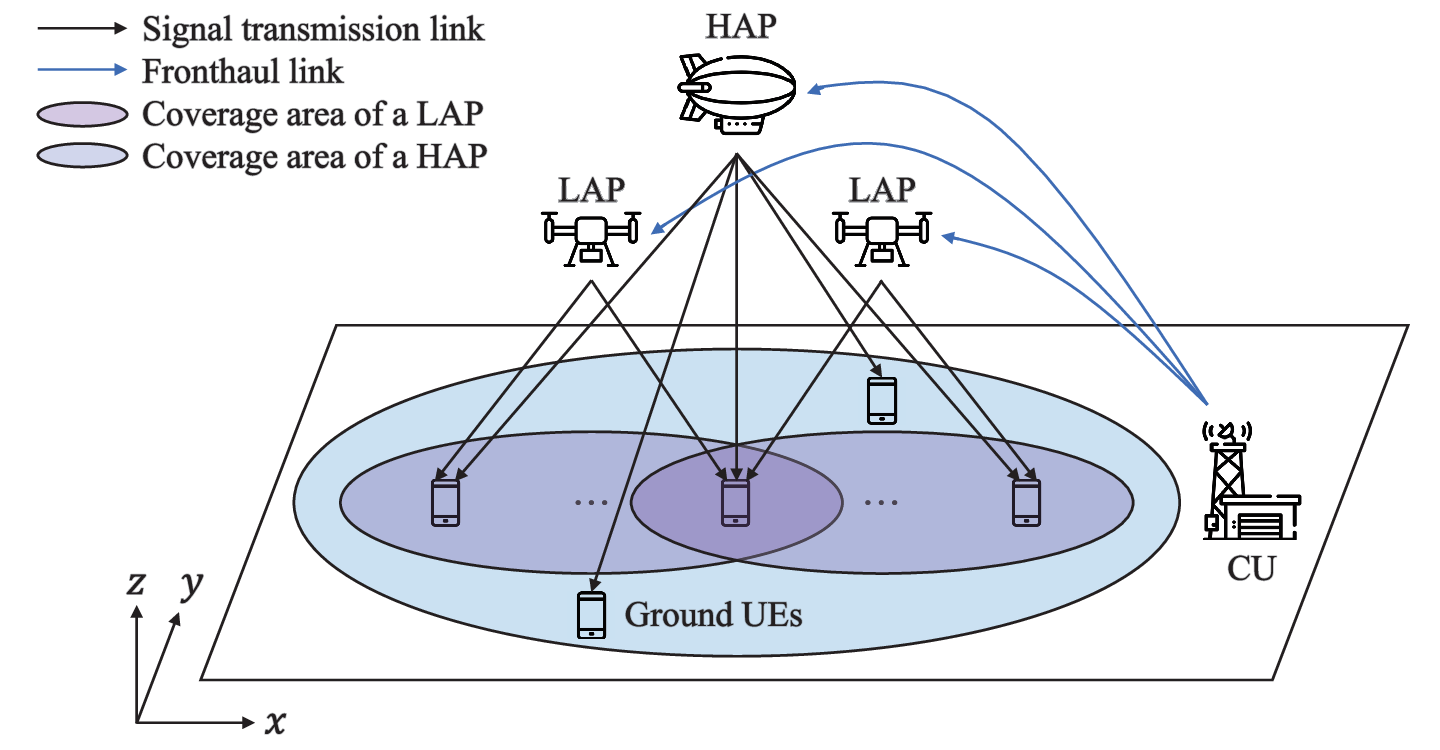}\vspace{-0.17cm}
    \caption{System model of hierarchical HAP-LAP networks.}
        \vspace{-0.6cm}
\label{fig:system}
\end{figure}

As shown in Fig. \ref{fig:system}, we consider the downlink cooperated HAP-LAP networks, consisting of CU, $K$ ground UEs and multiple aerial radio units (RUs) installed at HAP and LAPs. The set of UAVs is denoted as $\mathcal{U} = \left\{\mathcal{H},\mathcal{L}\right\}$, with $\mathcal{H}$ representing a HAP and $\mathcal{L}=\{1,\cdots,L\}$ representing the LAPs, respectively, while the set of $K$ ground UEs is defined as $\mathcal{K} = \left\{1,\cdots,K\right\}$. Here, a single HAP provides the globally wide coverage with long communication distance, while the LAPs are deployed to guarantee the relatively small end-to-end latency for compensating the high path-loss problem of the HAP. The consideration of a single cell managed by one HAP cooperated with multiple LAPs can be easily extended to the multiple cells with the interference managements \cite{karapantazis2005broadband} for scalability. All nodes of interest are assumed to be equipped with a single antenna. By means of C-RAN architecture \cite{hossain2019recent}, the HAP and LAPs are connected to and managed by the CU via constrained fronthaul links, where all baseband processing and overall system optimization are performed at the CU. Under the assumption of the orthogonal access of fronthaul links between CU and UAVs, the  CU compresses and forwards the messages for UEs to UAVs, which is well-known to provide the superior interference management, especially in dense networks \cite{kang2015fronthaul}. For the effective downlink communications of the multiple UEs, the RSMA scheme \cite{mao2022rate} is adopted, which makes it possible to efficiently reduce all levels of interference by using the rate-splitting and power allocation techniques. We assume that the channel state information (CSI) along with the placement information of nodes is assumed to be perfectly available at the CU, which can be extended to the imperfect CSI case by adopting the norm-bounded model \cite{jeong2015optimal}. 

Following \cite{jaafar2020downlink}, the air-to-ground channels are assumed to be dominated by line-of-sight (LoS) links. Accordingly, the channel gain from UAV $u\in\mathcal{U}$ to UE $k\in\mathcal{K}$ can be expressed as $g_{k,u}(\mathbf{q}_u^{U}) = \beta_u/\|\mathbf{q}_u^{U}-\mathbf{q}_k^{G}\|^2$, where $\beta_u$ is the received power at the reference distance of 1m, and $\mathbf{q}_u^{U}=\left(x_u^{U},y_u^{U},z_u^{U}\right)^{T}$ and $\mathbf{q}_k^{G}=\left(x_k^{G},y_k^{G},z_k^{G}\right)^{T}$ are the 3D placement of UAV $u$ and UE $k$. The received signal at UE $k$ is, therefore, given by
$y_k =\sum_{u\in \mathcal{U}_k}\sqrt{g_{k,u}(\mathbf{q}_u^{U})}x_u+n_k$, where $x_u$ is the transmit signal of UAV $u$, and $n_k \sim \mathcal{CN}(0, \sigma_k^2)$ is the additive white Gaussian noise (AWGN) at UE $k$ with $\sigma_k^2$ being the noise variance. Here, $\mathcal{U}_k$ is the set of UAVs that can be communicated with UE $k$, given as $\mathcal{U}_{k} = \{ u\in \mathcal{U} | \sqrt{(x_u^{U}-x_k^{G})^2+(y_u^{U}-y_k^{G})^2} \leq z_u\tan \theta_u \}$, with $\theta_u$ representing the UAV $u$'s coverage angle.

For the different flying height of HAP and LAPs, we take into account the different coverage area. We assume the larger coverage area of HAP than that of LAPs \cite{karapantazis2005broadband}, which is sufficiently enough that all $K$ UEs are within its coverage. Moreover, all UEs served by each UAV are supposed to communicate with the corresponding UAV, and the UEs served by multiple UAVs can therefore receive the superposition of the multiple received signals transmitted from the UAVs. For convenience, we define the set of UEs that are connected to UAV $u$ as $\mathcal{K}_u$, which is given as $\mathcal{K}_{u} = \{ k\in \mathcal{K} | \sqrt{(x_u^{U}-x_k^{G})^2+(y_u^{U}-y_k^{G})^2} \leq z_u\tan \theta_u \}$.

\vspace{-0.1cm}
\subsection{Rate-splitting and Fronthaul Compression}
For the RSMA technique \cite{mao2022rate}, the CU splits the UE $k$'s message $M_k$ into a private message $M^{p}_{k}$ and common messages $\{M^{c}_{k,u}\}_{u\in\mathcal{U}_k}$. The private message $M^{p}_{k}$, intended to be decoded by only UE $k$, is then encoded into a signal $s^{p}_{k}$ that follows a circularly symmetric complex Gaussian distribution $\mathcal{CN}(0,1)$. Similarly, the common messages $\{M^{c}_{k,u}\}_{k\in\mathcal{K}_u}$ are jointly encoded into $s^c_{u}\sim \mathcal{CN}(0,1)$, which can be decoded by the UEs belonging to $\mathcal{K}_u$ \cite{mao2022rate}. In this work, for the efficient message transfer and interference management within coverage, we propose that the HAP transmits both private and common messages, while the LAPs transmit common messages. As a result, the baseband signal for UAV $u$ at the CU can be expressed as $\tilde{x}_{u}=\sum_{k \in \mathcal{K}}\sqrt{P^p_k}s^p_k+\sqrt{P^c_u}s^c_u$, for $u\in\mathcal{H}$ and 
$\tilde{x}_{u}=\sqrt{P^c_u}s^c_u$, for $u\in\mathcal{L}$, where $P^p_{k}$ and $P^c_{u}$ represent the transmit power for the private signal $s^p_{k}$ and for the common signal $s^c_{u}$, respectively. The CU compresses and forwards the baseband signal $\tilde{x}_{u}$ to UAV $u$ via the $u$-th fronthaul link with the limited capacity $C_u$. In order to facilitate analysis and design, we adopt the standard random coding approach \cite{kang2015fronthaul} of rate-distortion theory to model the effect of quantization by means of an additive quantization noise. Consequently, we can express the quantized signal $x_u$ as $x_{u}= \tilde{x}_{u}+\omega_{u}$, where quantization noise is defined as $\omega_{u}\sim \mathcal{CN}(0,\sigma_{\omega_{u}}^2)$, with $\sigma_{\omega_{u}}^2$ representing the variance of quantization noise. Furthermore, the rate-distortion theory \cite{kang2015fronthaul} enables us to quantify the required fronthaul rate between CU and UAV $u$ as $R_{u}(\mathbf {P}_u,\sigma_{\omega_{u}}^2)$, which is constrained by  
\begin{align}
R_{u}(\mathbf {P}_u,\sigma_{\omega_{u}}^2)=\log_2\left(1+\sigma^2_{\tilde{x}_{u}}(\mathbf{P}_u)/\sigma^2_{\omega_{u}}\right)\leq C_{u},u\in \mathcal {U},
\label{Backhaul capacity constraint}
\end{align}
where $\mathbf{P}_u$ is the set of transmit powers for UAV $u$ and $\sigma^2_{\tilde{x}_{u}}(\mathbf{P}_u)=\mathbb{E}[\tilde{x}_{u}\tilde{x}_{u}^{H}]$ is the variance of $\tilde{x}_u$. Due to the operational capability of UAVs, we have the constraint as 
\begin{align}
P_{u}(\mathbf {P}_u,\sigma^2_{\omega_{u}})=\sigma^2_{\tilde{x}_{u}}(\mathbf{P}_u)+\sigma^2_{\omega_{u}}\leq P^{\textrm{max}}_{u},u\in \mathcal {U},
\label{Transmit power constraint}
\end{align}
where $P_u^{\textrm{max}}$ denotes the maximum transmit power of UAV $u$.

Due to the coexistence of HAP and LAPs, each UE experiences the interference, which can be mitigated by successive interference cancellation (SIC). In particular, the UE $k$ decodes the common signals $\{s^{c}_{u}\}_{u \in \mathcal{U}_k}$ by treating the interference as noise. Similarly, the private messages of other UEs can be treated in decoding as interference. To describe the decoding order of common messages using SIC at UE $k$, we introduce a permutation order $\pi_{k} :  \mathcal{U}_k  \rightarrow \left\{1,\cdots,U_{k}\right\}$, where $U_k=|\mathcal{U}_k|$ is the number of UAVs associated with UE $k$. Based on the definition of $\pi_{k}$ and with the perfect SIC assumption, the achievable rate of the common messages is constrained as 
\vspace{-0.1cm}
\begin{align}
\sum_{k\in \mathcal{K}_{u}}R^{c}_{k,u}\leq \min_{k\in \mathcal{K}_{u}}f^{c}_{k,u}(\mathbf{P},\bm{\sigma}^2,\mathbf{q}),u\in \mathcal {U},
\label{Achievable rate constraint}
\end{align}
where $\mathbf{P}=\{\mathbf{P}_u\}_{u\in\mathcal{U}}$, $\bm{\sigma}^2=\{\sigma^2_{\omega_u}\}_{u\in\mathcal{U}}$, $\mathbf{q}= \left\{ \mathbf{q}_u^{U}\right\}_{u\in \mathcal{U}}$ and $f^{c}_{k,u} (\mathbf{P},\bm{\sigma}^2,\mathbf{q})$ is defined as $f^{c}_{k,u}(\mathbf{P},\bm{\sigma}^2,\mathbf{q})=\log_2(1+P^{c}_{u}g_{k,u}(\mathbf{q}_u^{U})/\nu^{c}_{k,u}(\mathbf{P},\bm{\sigma}^2,\mathbf{q}))$, with
\vspace{-0.1cm}
\begin{align}
&\nu^{c}_{k,u}(\mathbf{P},\bm{\sigma}^2,\mathbf{q})=\underbrace{\sum_{j=\pi_{k}(u)+1}^{U_k}P^{c}_{\pi_k^{-1}(j)}g_{k,\pi_k^{-1}(j)}\left(\mathbf{q}_{\pi_k^{-1}(j)}^{U}\right)}_{\textrm{Uncanceled common messages}}
\nonumber\\
&+\underbrace{\sum_{j\in\mathcal{K}}P^{p}_{j}g_{k,\mathcal{H}}\left(\mathbf{q}_\mathcal{H}^{U}\right)}_{\textrm{Private messages for all UEs}}+\underbrace{\sum_{i\in\mathcal{U}_k}\sigma_{\omega_{i}}^2g_{k,i}\left(\mathbf{q}_i^{U}\right)}_{\textrm{Quantization noise}}+\underbrace{\sigma_{k}^2}_{\textrm{AWGN}},
\end{align}
and $\pi_k^{-1}$ indicating the inverse permutation of $\pi_k$.

After the common signals $\{s^{c}_{u}\}_{u \in \mathcal{U}_k}$ are perfectly canceled at UE $k$, UE $k$ decodes its corresponding private signal $s^{p}_{k}$ from the remaining received signal. With the SIC decoding process, the achievable rate of the private message of UE $k$ is given by $R^{p}_{k}(\mathbf{P},\bm{\sigma}^2,\mathbf{q})= \log_2(1+P^{p}_{k}g_{k,\mathcal{H}}(\mathbf{q}_\mathcal{H}^{U})/\nu^{p}_{k}(\mathbf{P},\bm{\sigma}^2,\mathbf{q}))$, with
\begin{align}
\nu^{p}_{k}(\mathbf{P}\!,\bm{\sigma}^2\!,\mathbf{q})\!=\!\!\!\!\!\underbrace{\sum_{j\in\mathcal{K}\setminus\left\{k\right\}}\!P^{p}_{j}\!g_{k,\mathcal{H}}\!\left(\mathbf{q}_\mathcal{H}^{U}\right)}_{\textrm{Private messages from other UEs}}\!\!\!\!+\!\!\underbrace{\sum_{i\in\mathcal{U}_k}\sigma_{\omega_{i}}^2g_{k,i}\left(\mathbf{q}_i^{U}\right)}_{\textrm{Quantization noise}}\!+\!\!\!\underbrace{\sigma_{k}^2}_{\textrm{AWGN}}.
\end{align}
The achievable rate of UE $k$ can be calculated by the sum of common and private messages, which needs to satisfy the minimum rate requirement: 
\vspace{-0.1cm}
\begin{align}
R^{\textrm{th}}_{k} \leq R^{p}_{k}(\mathbf{P},\bm{\sigma}^2,\mathbf{q})+\sum_{u\in\mathcal{U}_{k}}R^{c}_{k,u},k\in \mathcal {K},\label{QoS constraint}
\end{align}
with $R_k^{\textrm{th}}$ representing the minimum required rate at UE $k$.

\vspace{-0.2cm}
\section{Sum-rate Maximization of Hierarchical HAP-LAP Networks}
In this paper, our goal is to maximize the sum-rate by jointly optimizing the rate-splitting variables of $\mathbf{R} = \{ R^{c}_{k,u}\}_{u\in\mathcal{U},k\in \mathcal{K}_u}$, the transmit powers of $\mathbf{P}$, the variances of quantization noise $\bm{\sigma}^2$, and the UAV placements of $\mathbf{q}$. The problem is then formulated as
\vspace{-0.1cm}
\begin{subequations}
\label{main_problem}
\begin{align}
\mathrm{(P1)\,}&\mathrm{:}\max\limits_{\mathbf {R},\mathbf {P},\bm{\sigma}^2,\mathbf{q}}\sum _{k\in \mathcal {K}}\left({R^{p}_{k}(\mathbf{P},\bm{\sigma}^2,\mathbf{q})+\sum_{u\in\mathcal{U}_k}R^{c}_{k,u}}\right) \label{P1 objective}\\
\mathrm{s}.\mathrm{t}.~&d_{\min }^2\leq{\left\Vert {{{\mathbf {q}}_u^{U}} - {{\mathbf {q}}_{u'}^{U}}} \right\Vert}^2 ,\,u \neq u' \textrm{ and } u,u' \in \mathcal{U},\label{UAV safety constraint}\\
&x^\textrm{min}_u\leq x_u^{U} \leq x^\textrm{max}_u\,, y^\textrm{min}_u \leq y_u^{U} \leq y^\textrm{max}_u,u\in \mathcal {U} \label{UAV placement constraint1},\\
&z^\textrm{min}_u \leq z_u^{U} \leq z^\textrm{max}_u,u\in \mathcal {U} \label{UAV placement constraint2},\\
&\eqref{Backhaul capacity constraint},\,\eqref{Transmit power constraint},\,\eqref{Achievable rate constraint}\,\,\text{and}\,\,\eqref{QoS constraint}\label{P1 constraint},
\end{align}
\label{P1}
\end{subequations}
where \eqref{UAV safety constraint} represents the UAV's safety constraint that ensures the avoidance of collisions among different UAVs, and \eqref{UAV placement constraint1} and \eqref{UAV placement constraint2} ensure that all UAVs are placed with in the predetermined area.
The problem P1 is non-convex due to its non-concave objective function \eqref{P1 objective} and the non-convex constraints \eqref{Backhaul capacity constraint}, \eqref{Achievable rate constraint}, \eqref{QoS constraint} and \eqref{UAV safety constraint}. To deal with the non-concavity of \eqref{P1 objective}, we introduce the slack variable $\boldsymbol{\eta}^p=\{\eta^p_k\}_{k\in\mathcal{K}}$, by which the problem \eqref{P1} is reformulated as
\vspace{-0.1cm}
\begin{subequations} \label{P2}
\begin{align}
\mathrm{(P2)\,}&\mathrm{:}\max\limits_{\mathbf {R},\mathbf {P},\bm{\sigma}^2,\mathbf{q},\boldsymbol{\eta}^p }\sum _{k\in \mathcal {K}}\left({ \eta^p_{k}+\sum_{u\in\mathcal{U}_{k}}R^{c}_{k,u}}\right) \label{P2_objfuc}\\ 
\mathrm{s}.\mathrm{t}.~&\eta^p_{k}\leq R^{p}_{k}(\mathbf{P},\bm{\sigma}^2,\mathbf{q}), k\in \mathcal {K} \label{Achievable private rate constraint},\\
&\eqref{UAV safety constraint}-\eqref{P1 constraint}.
\end{align}
\end{subequations} 
For solving \eqref{P2}, we develop the AO-based algorithm to iteratively solve two sub-problems, whose details are given in the following. 

\vspace{-0.3cm}
\subsection{Optimization of UAV Placement}
Given $\{\mathbf {R},\mathbf {P},\bm{\sigma}^2\}$ in P2, the optimal solution of $(R^c_{k,u})^*$ is obtained with the equality of \eqref{Achievable rate constraint} to maximize the achievable rates of common messages, which allows to transform \eqref{P2_objfuc} into $\sum _{k\in \mathcal {K}}\eta^p_{k} +\sum_{u\in\mathcal{U}}\min_{k\in\mathcal{K}_u}f^c_{k,u}(\mathbf{P},\bm{\sigma}^2,\mathbf{q})$.
By introducing the slack variable $\boldsymbol{\eta}^c=\{\eta^c_u\}_{u\in\mathcal{U}}$, we have the more tractable problem
\vspace{-0.1cm}
\begin{subequations}
\begin{align}
\mathrm{(P3)\,}&\mathrm{:}\max\limits_{\mathbf{q},\boldsymbol{\eta}^p,\boldsymbol{\eta}^c}\sum _{k\in \mathcal {K}}\eta^p_{k} +\sum_{u\in\mathcal{U}}\eta^c_{u}\\ 
\qquad \mathrm{s}.\mathrm{t}.~&\eta^c_u\leq f_{k,u}^{c}\left ({\mathbf {P},\bm{\sigma}^2,\mathbf{q}}\right),u\in \mathcal {U},k\in \mathcal {K}_{u},\label{Achievable common rate constraint}\\
&\eqref{Achievable rate constraint},\,\eqref{QoS constraint},\,\eqref{UAV safety constraint}-\eqref{UAV placement constraint2}\,\,\text{and}\,\,\eqref{Achievable private rate constraint}\label{P3 constraint}.
\end{align}
\end{subequations}
To tackle with the non-convex constraints, we adopt the SCA technique \cite{boyd2004convex}, which involves replacing the non-convex parts with the tractable convex approximates in each iteration. Following \cite{boyd2004convex}, we firstly rewrite the function $f^c_{k,u}(\mathbf{P},\bm{\sigma}^2,\mathbf{q})$ in \eqref{Achievable rate constraint} and \eqref{Achievable common rate constraint} as
\vspace{-0.1cm}
\begin{align}
\!\!\!\!f^c_{k,u}(\mathbf{P},\bm{\sigma}^2,\mathbf{q})\!=\!\hat f^c_{k,u}(\mathbf{P},\bm{\sigma}^2,\mathbf{q})\!-\!\log_2\left(\nu^c_{k,u}\left(\mathbf{P},\bm{\sigma}^2,\mathbf{q}\right)\right)\!,
\end{align}
where we define 
\vspace{-0.1cm}
\begin{align}
\hat{f}^c_{k,u}(\mathbf{P},\bm{\sigma}^2,\mathbf{q}) = \log _{2}\left ({\sum _{i\in\mathcal{U}_k}\frac {\beta_{i} A_{k,u,i}(\mathbf{P},\bm{\sigma}^2)}{\|\mathbf{q}_i^{U}-\mathbf{q}_k^{G}\|^{2}}+\sigma_{k} ^{2}}\right)
\end{align}
and 
\vspace{-0.1cm}
\begin{align}
&A_{k,u,i}(\mathbf{P},\bm{\sigma}^2)\nonumber\\
&\!\!=\!\begin{cases}
\sigma_{\omega_{i}}^2,\qquad\qquad\, i(\ne \mathcal{H})\!\in\!\mathcal{U}_{k} \!\cap\! \left\{ \pi_k\left(u\right) \!>\!\pi_k\left(i\right)\right\},\\
\sigma_{\omega_{i}}^2\!+\!P^c_{i},\qquad\, i(\ne \mathcal{H})\!\in\!\mathcal{U}_{k} \!\cap\! \left\{ \pi_k\left(u\right) \!\leq\!\pi_k\left(i\right)\right\},\\
\sigma_{\omega_{i}}^2\!+\!\sum_{j\in\mathcal{K}}P^p_{j},\qquad\,\, i\!\in\!\mathcal{H} \!\cap\! \left\{ \pi_k\left(u\right) \!>\!\pi_k\left(i\right)\right\},\\
\sigma_{\omega_{i}}^2\!+\!P^c_{i}\!+\!\sum_{j\in\mathcal{K}}P^p_{j},\,\,i\!\in\!\mathcal{H} \!\cap\! \left\{ \pi_k\left(u\right) \!\leq\!\pi_k\left(i\right)\right\}.\\
\end{cases}
\label{A}
\end{align}
Similarly, by applying the SCA to $\hat f^c_{k,u}(\mathbf{P},\bm{\sigma}^2,\mathbf{q})$ and introducing the slack variable $\mathbf{S}=\{S_{k,i}\}_{k\in\mathcal{K},i\in\mathcal{U}_k}$ to satisfy $S_{k,i} \leq \|\mathbf{q}_i^{U}-\mathbf{q}_k^{G}\|^{2},\,k\in \mathcal {K},i\in \mathcal {U}_k\label{slack S}$, we can obtain the concave lower bound of $f^c_{k,u}(\mathbf{P},\bm{\sigma}^2,\mathbf{q})$ as
\vspace{-0.1cm}
\begin{align}
\!f^c_{k,u}(\mathbf{P},\!\bm{\sigma}^2,\!\mathbf{q})&\geq f_{k,u}^{c,\textrm{lb}}(\mathbf{P},\!\bm{\sigma}^2,\!\mathbf{q},\!\mathbf{S})\nonumber\\
&\triangleq \!\hat{f}_{k,u}^{c,\textrm{lb}}(\mathbf{P},\!\bm{\sigma}^2,\!\mathbf{q})\!-\!\log _{2}\!\left (\check{\nu}^c_{k,u}(\mathbf{P},\!\bm{\sigma}^2,\!\mathbf{q},\!\mathbf{S})\right)\!,\!\label{concave_lower_f_c}\end{align}
which is obtained by the first-order Taylor approximation, where
\vspace{-0.1cm}
\begin{align}
&\hat f^{c,\textrm{lb}}_{k,u}(\mathbf{P},\bm{\sigma}^2,\mathbf{q})=\! \log _{2}\left ({\sum _{j\in\mathcal{U}_k}\frac {\beta_{j} A_{k,u,j}(\mathbf{P},\bm{\sigma}^2)}{\|\mathbf{q}_j^{U,(r)}-\mathbf{q}_k^{G}\|^{2}}+\sigma_{k} ^{2}}\right) \nonumber\\
&\!\!-\!\!\sum _{i\in\mathcal{U}_k}\!\!\frac { \frac {\beta_{i} A_{k,u,i}(\!\mathbf{P}\!,\!\bm{\sigma}^2)}{\|\mathbf{q}_i^{U,(r)}-\mathbf{q}_k^{G}\|^{4}}\log _{2}(e)}{\sum _{j\in\mathcal{U}_k} \!\!\frac {\beta_{j} A_{k,u,j}(\mathbf{P},\bm{\sigma}^2)\!\!}{\!\|\mathbf{q}_j^{U,(r)}-\mathbf{q}_k^{G}\|^{2}} \!+\! \sigma_k ^{2} }\!\!\left (\!{\|\mathbf{q}_i^{U}\!\!-\!\mathbf{q}_k^{G}\|^{2} \!-\!\|\mathbf{q}_i^{U,(r)}\!\!-\!\mathbf{q}_k^{G}\|^{2}}\right ).\label{lower bound common}
\end{align}
In same manner, for handling the non-convex constraints \eqref{QoS constraint} and \eqref{Achievable private rate constraint}, we can derive the concave lower bound of $R^{p}_{k}(\mathbf{P},\bm{\sigma}^2,\mathbf{q})$, denoted as $R_{k}^{p,\textrm{lb}}\left(\mathbf{P},\bm{\sigma}^2,\mathbf{q},\mathbf{S}\right)$. In order to handle with the non-convexity of $\eqref{UAV safety constraint}$, we apply the first-order Taylor expansion \cite{boyd2004convex} at the given points $\mathbf {q}_u^{U,(r)}$ and $\mathbf {q}_{u'}^{U,(r)}$, which yields, for $u \neq u'$ and $u,u' \in \mathcal{U}$, $\Vert {{{\mathbf {q}}_u^U} - {{\mathbf {q}}_{u'}^U}}\Vert^2\geq2(\mathbf {q}_{u}^{U,(r)}-\mathbf {q}_{u'}^{U,(r)})^{T} ( \mathbf {q}_{u}^U-\mathbf {q}_{u'}^U)-\|\mathbf {q}_{u}^{U,(r)}-\mathbf {q}_{u'}^{U,(r)}\|^{2}\triangleq \psi_{u,u'}(\mathbf{q})$, and, for $k\in \mathcal {K}$ and $i\in \mathcal {U}_k$, $\|\mathbf{q}_i^{U}-\mathbf{q}_k^{G}\|^{2}\geq2(\mathbf {q}_{i}^{U,(r)}-\mathbf {q}_{k}^{G})^{T} ( \mathbf {q}_{i}^U-\mathbf {q}_{i}^{U,\left(r\right)})+\|\mathbf {q}_{i}^{U,(r)}-\mathbf {q}_{k}^{G}\|^{2}\triangleq \tau_{k,i}(\mathbf{q})$.
Finally, we can have the convex problem for UAV placement 
\vspace{-0.1cm}
\begin{subequations}
\begin{align}
\mathrm{(P4)\,}&\mathrm{:}\max\limits_{\mathbf{q},\boldsymbol{\eta}^p,\boldsymbol{\eta}^c,\mathbf{S}}\sum _{k\in \mathcal {K}}\eta^p_{k} +\sum_{u\in\mathcal{U}}\eta^c_{u}\\ 
\mathrm{s}.\mathrm{t}.~&\eta^p_{k}\leq R_{k}^{p,\textrm{lb}}\left(\mathbf{P},\bm{\sigma}^2,\mathbf{q},\mathbf{S}\right), k\in \mathcal {K},\\
&\eta^c_u\leq f^{c,\textrm{lb}}_{k,u}\left ({\mathbf {P},\bm{\sigma}^2,\mathbf{q}},\mathbf{S}\right),u\in \mathcal {U},k\in \mathcal {K}_{u},\\
&\sum _{k\in \mathcal {K}_{u}}R^{c}_{k,u}\!\leq\! f^{c,\textrm{lb}}_{k,u}\left ({\mathbf {P},\bm{\sigma}^2,\mathbf{q}},\mathbf{S}\right),u\in \mathcal {U},k\in \mathcal {K}_{u},\\
&R^{\textrm{th}}_{k}\leq R_{k}^{p,\textrm{lb}}\left(\mathbf{P},\bm{\sigma}^2,\mathbf{q},\mathbf{S}\right)+\sum_{u\in\mathcal{U}_{k}}R^{c}_{k,u},k\in \mathcal {K},\\
&d_{\min }^2\leq \psi_{u,u'}(\mathbf{q}),\,u \neq u' \textrm{ and } u,u' \in \mathcal{U},\\
&S_{k,i} \leq \tau_{k,i}(\mathbf{q}),\,k\in \mathcal {K},i\in \mathcal {U}_k,\\
&\eqref{UAV placement constraint1}\,\,\text{and}\,\,\eqref{UAV placement constraint2},
\end{align}
\end{subequations}
which can be solved by strong duality theorem or CVX \cite{grant2014cvx}. 

\vspace{-0.1cm}
\subsection{Optimization of RSMA and Fronthaul Design}
For the fixed $\mathbf{q}$, the problem P2 can be simplified into
\vspace{-0.1cm}
\begin{subequations}
\begin{align}
\mathrm{(P5)\,}&\mathrm{:}\max\limits_{\mathbf {R},\mathbf {P},\bm{\sigma}^2,\boldsymbol{\eta}^p }\sum _{k\in \mathcal {K}}\left({ \eta^p_{k}+\sum_{u\in\mathcal{U}_{k}}R^{c}_{k,u}}\right)\\
\mathrm{s}.\mathrm{t}.~&\eqref{Backhaul capacity constraint},\,\eqref{Transmit power constraint},\,\eqref{Achievable rate constraint},\,\eqref{QoS constraint}\,\,\text{and}\,\,\eqref{Achievable private rate constraint}.
\end{align}
\end{subequations}
To tackle with the non-convexity of P5, we adopt the WMMSE method \cite{yu2019efficient} that facilitates the conversion of non-convex optimization problem into convex programming by appropriately adjusting the receive equalizers and weights to minimize the MSE. Based on WMMSE method \cite{yu2019efficient}, we can have $f^c_{k,u}(\mathbf {P},\bm{\sigma}^2,\mathbf{q})\geq\widetilde {f}^{c}_{k,u}({\mathbf {P},\bm{\sigma}^2,\mathbf{q},e^c_{k,u},w^c_{k,u}})\triangleq\log _{2}w^c_{k,u}+(1-w^c_{k,u}\varepsilon^c_{k,u}({\mathbf {P},\bm{\sigma}^2,\mathbf{q},e^c_{k,u}}))/\ln 2$ in \eqref{Achievable rate constraint} and $R^p_{k}(\mathbf {P},\bm{\sigma}^2,\mathbf{q})\geq\widetilde {R}^p_{k} ({\mathbf {P},\bm{\sigma}^2,\mathbf{q},e^p_{k},w^p_{k}})
\triangleq\log _{2}w^p_{k}+({1-w^p_{k}\varepsilon^p_{k}({\mathbf {P},\bm{\sigma}^2,\mathbf{q},e^p_{k}})})/\ln 2$ in \eqref{QoS constraint} and \eqref{Achievable private rate constraint}, where $e^c_{k,u}$ and $e^p_{k}$ represent any receive equalizers, while $w^c_{k,u}$ and $w^p_{k}$ are non-negative weights associated with MSE.
Additionally, the MSE $\varepsilon^c_{k,u}({\mathbf {P},\bm{\sigma}^2,\mathbf{q},e^c_{k,u}})$ of common signal and the MSE $\varepsilon^p_{k}({\mathbf {P},\bm{\sigma}^2,\mathbf{q},e^p_{k}})$ of private signal are defined as $\varepsilon^c_{k,u}({\mathbf {P},\bm{\sigma}^2,\mathbf{q},e^c_{k,u}})
=(e^c_{k,u})^2\nu^{c}_{k,u}(\mathbf{P},\bm{\sigma}^2,\mathbf{q})+(1-e^c_{k,u}\sqrt{P^c_{u}g_{k,u}(\mathbf{q}_u^{U})})^2$ and $\varepsilon^p_{k} ({\mathbf {P},\bm{\sigma}^2,\mathbf{q},e^p_{k}})=(e^p_k)^2\nu^{p}_{k}(\mathbf{P},\bm{\sigma}^2,\mathbf{q})+(1-e^p_k\sqrt{P^p_{k}g_{k,\mathcal{H}}(\mathbf{q}_\mathcal{H}^{U})})^2$, respectively.
Note that the lower bounds $f^c_{k,u}(\mathbf {P},\bm{\sigma}^2,\mathbf{q})$ and $\varepsilon^p_{k}({\mathbf {P},\bm{\sigma}^2,\mathbf{q},e^p_{k}})$ are sufficiently tight when the equalizers $\{e^c_{k,u},e^p_{k}\}$ and the weights $\{w^c_{k,u},w^p_{k}\}$ satisfy the following conditions \cite{yu2019efficient}: 
\begin{align}
&\!\!\!\!\!e^c_{k,u}\!\!=\!\!\sqrt{P^c_{u}g_{k,u}\left(\mathbf{q}_u^{U}\right)}/\left(\nu^c _{k,u}\left(\mathbf {P},\bm{\sigma}^2,\mathbf{q}\right)+P^c_{u}g_{k,u}\left(\mathbf{q}_u^{U}\right)\right),\\
&\!\!\!\!\!e^p_{k}\!=\!\sqrt{P^p_{k}g_{k,\mathcal{H}}\left(\mathbf{q}_\mathcal{H}^{U}\right)}/\left(\nu^p _{k}\left(\mathbf {P},\bm{\sigma}^2,\mathbf{q}\right)+ P^p_{k}g_{k,\mathcal{H}}\left(\mathbf{q}_\mathcal{H}^{U}\right)\right),\\
&\!\!\!\!\!w^c_{k,u}\!\!=\!\!1\!/\varepsilon^c_{k,u}\!\!\left({\mathbf {P},\bm{\sigma}^2\!,\mathbf{q},e^c_{k,u}}\right) \,\,\textrm{and}\,\,w^p_{k}\!\!=\!\!1\!/\varepsilon^p_{k}\!\left({\mathbf {P},\bm{\sigma}^2\!,\mathbf{q},e^p_{k}}\right). 
\end{align}
To deal with the non-convex function $R_{u}(\mathbf {P}_u,\sigma_{\omega_{u}}^2)$ in $\eqref{Backhaul capacity constraint}$, we utilize the inequality derived from the concavity of the $\log\left(·\right)$ function as $R_{u}(\mathbf {P}_u,\sigma_{\omega_{u}}^2)
\leq\widetilde {R}_{u} (\mathbf{P}_u,\sigma_{\omega_{u}}^2,\Sigma _{u})\triangleq\log _{2}(\Sigma_{u})+((\sigma^2_{\tilde{x}_{u}}(\mathbf{P}_u)+\sigma_{\omega_{u}}^2)/\Sigma_{u}-1)/\ln 2-\log _{2}\left(\sigma_{\omega_{u}}^2\right)$, which holds for an arbitrary positive real number $\boldsymbol{\Sigma }_{u}$, and whose equality is achieved if and only if $\boldsymbol{\Sigma }_{u}=\sigma^2_{\tilde{x}_{u}}(\mathbf{P}_u)+\sigma_{\omega_{u}}^2$.

As a result, we can transform the problem P5 into
\begin{subequations}
\begin{align}
\mathrm{(P6)\,}&\mathrm{:}\max\limits_{\mathbf {R},\mathbf {P},\bm{\sigma}^2,\boldsymbol{\eta}^p,\mathbf{e},\mathbf{w},\boldsymbol{\Sigma }}~\sum _{k\in \mathcal {K}}\left({\eta^p_{k}+\sum_{u\in\mathcal{U}_k}R^c_{k,u}}\right)\\ 
\mathrm{s}.\mathrm{t}.~&\eta^p_{k}  \leq \widetilde {R}^p_{k}\left ({\mathbf {P},\bm{\sigma}^2,\mathbf{q},e^p_{k},w^p_{k}}\right),k\in \mathcal {K},\\
&\sum _{k\in \mathcal {K}_{u}}R^c_{k,u}\!\leq\! \widetilde {f}^c_{k,u}\left ({\mathbf {P},\bm{\sigma}^2,\mathbf{q},e^c_{k,u},w^c_{k,u}}\right),u\in \mathcal {U},\!k\in \mathcal {K}_{u},\\
&R^{\textrm{th}}_{k}\leq \widetilde {R}^p_{k}\left ({\mathbf {P},\bm{\sigma}^2,\mathbf{q},e^p_{k},w^p_{k}}\right)\!+\!\sum_{u\in\mathcal{U}_{k}}R^c_{k,u},k\in \mathcal {K},\\
&\widetilde {R}_{u}\left (\mathbf {P}_u,\bm{\sigma}^2,\Sigma _{u}\right)\leq C_{u},u\in \mathcal {U},\\
&\eqref{Transmit power constraint}.
\end{align}
\end{subequations}
where $\mathbf{e} = \{ e^p_{k}\}_{k\in \mathcal{K}} \cup \{e^c_{k,u}\}_{u\in\mathcal{U},k\in \mathcal{K}_{u}}$,
$\mathbf{w} = \left\{ w_{p,k}\right\}_{k\in \mathcal{K}} \cup \left\{w_{c,k,u}\right\}_{u\in\mathcal{U},k\in \mathcal{K}_{u}}$ and $\boldsymbol{\Sigma } = \left\{ \boldsymbol{\Sigma }_{u}\right\}_{u\in\mathcal{U}}$. The problem P6 is convex for the fixed $\mathbf{e}, \mathbf{w}$ and $\boldsymbol{\Sigma}$, which can be solved by convex optimization solvers, e.g., CVX \cite{grant2014cvx}. By using P4 and P6, we can obtain the local optimal solution of P1 by using AO method, whose details are summarized in Algorithm 1. 

\vspace{-0.15cm}
\subsection{Convergence and Complexity Analysis}
\begin{algorithm}[t]
\small
\caption{Sum-rate Maximization Algorithm}
\begin{algorithmic}[1]
\STATE $\textbf{Initialization :}$ $r\leftarrow 0$, $(\mathbf {R}^{(r)},\mathbf {P}^{(r)},\bm{\sigma}^{2,(r)},\mathbf{q}^{(r)},R_{\textrm{sum}}^{(r)})$;
\REPEAT
\STATE $r\leftarrow r+1$;
\STATE Given $(\mathbf {R}^{(r-1)},\mathbf {P}^{(r-1)},\bm{\sigma}^{2,(r-1)})$, solve P4 using SCA method and get $\mathbf {q}^{(r)}$;
\STATE Adjust the sets $\{\mathcal{K}_u$, $\mathcal{U}_k\}$ and $\pi_k$ according to $\mathbf {q}^{(r)}$;
\STATE Given $\mathbf {q}^{(r)}$, solve P6 using WMMSE method to obtain $(\mathbf {R}^{(r)},\mathbf {P}^{(r)},\bm{\sigma}^{2,(r)},R_{\textrm{sum}}^{(r)})$ and update $\pi_k$;
\UNTIL{$\left|R_{\textrm{sum}}^{(r)}-R_{\textrm{sum}}^{(r-1)}\right|\leq\epsilon$};
\STATE $\textbf{Return }(\mathbf {R}^{(r)},\mathbf {P}^{(r)},\bm{\sigma}^{2,(r)},\mathbf{q}^{(r)},R_{\textrm{sum}}^{(r)})$
\end{algorithmic}
\end{algorithm}
For solving the problem P1, we propose the AO-based Algorithm 1, which guarantees that the sum-rate monotonically increases with each iteration. Consequently, the proposed algorithm ensures the convergence by repeating until $|R_{\textrm{sum}}^{(r)}-R_{\textrm{sum}}^{(r-1)}|\leq\epsilon$ is satisfied, where $r$ is the iteration index and $\epsilon \ll 1$ is the convergence condition.

The complexity of Algorithm 1 is mainly determined by the sub-problems P4 and P6. To address the non-convex UAV placment optimization, we convert it into the convex problem P4 with $N_1$ constraints and $N_2$ variables and its complexity is given as $\mathcal{O}((LK)^{1.5}\log_2(1/\epsilon))$ \cite{grant2014cvx}, where $N_1$ and $N_2$ are defined as $N_1=(L+1)(3K+L+4)+2K$ and $N_2=(L+1)(K+4)+K$, respectively. Also, P6 involves with $N_3=(L+1)(K+2)+2K$ variables, which yields the complexity of $\mathcal{O}(Y_1(LK)^{3})$, where $Y_1$ is the number of iterations for P6 \cite{boyd2004convex}. Finally, the complexity of Algorithm 1 can be calculated by  $\mathcal{O}(Y_2(LK)^{1.5}\log_2(1/\epsilon)+Y_2Y_1(LK)^{3})$, where $Y_2$ denotes the number of iterations in Algorithm 1.

\vspace{-0.2cm}
\begin{table}[t]
\centering
\caption{Simulation parameters}
\vspace{-0.1cm}
\begin{tabular}[t]{ccl}
\hline
\hline
&HAP&LAP\\
\hline
Maximum UAV altitude $(z_u^{\textrm{min}})$ & 17 km \cite{karapantazis2005broadband}& 2 km\\
Minimum UAV altitude $(z_u^{\textrm{max}})$ & 22 km \cite{karapantazis2005broadband}& 3 km\\
Maximum transmission power $(P_u^{\textrm{max}})$& 10 W& 0.5 W\\
Channel power gain at the reference distance $(\beta_u)$& -10 dB& -20 dB\\
UAV coverage angle $(\theta_u)$ &\multicolumn{2}{c} {$\pi/4$}\\
Minimum inter UAV distance $(d_{\textrm{min}})$ &\multicolumn{2}{c} {2 km}\\
Receiver noise power $(\sigma^2_k)$ &\multicolumn{2}{c} {-100 dBm}\\
Minimum required rate $(R^{\textrm{th}}_{k})$ &\multicolumn{2}{c} {0.1 bps/Hz}\\
\hline
\hline
\end{tabular}
  \vspace{-0.4cm}
\label{uav parameter setting}
\end{table}

\vspace{-0.1cm}       
\section{Numerical results}
In this section, we provide the numerical results to evaluate the proposed Algorithm 1. In simulations, the UEs are randomly distributed within a $10 \times 10 \, \textrm{km}^2$ square area and we set the initial altitude of the UAV to the maximum allowed value. For the implementation of RSMA, the UEs decode signals sequentially, starting with the one the highest power \cite{yu2019efficient}. We assume that the total fronthaul capacity $C_{T}$ is allocated to HAP and LAPs, where $C_{\mathcal{H}}=2C_{u}$, for $u\in\mathcal{L}$. The remaining parameters are set as described in Table \ref{uav parameter setting} following by \cite{kang2023cooperative}, \cite{karapantazis2005broadband}. For performance comparisons, we consider three types of benchmark schemes: \lowercase\expandafter{\romannumeral1}) $\textit{HAP-LAP w/o plc opt.}$, where HAP-LAP networks are designed by Algorithm 1 without placement optimization; \lowercase\expandafter{\romannumeral2}) $\textit{HAP only}$, where the HAP exists only with Algorithm 1; \lowercase\expandafter{\romannumeral3}) $\textit{HAP-LAP w/o pwr opt.}$, where the equal power allocation is applied to all messages in Algorithm 1.

\begin{figure}[t]
  \centering
   \includegraphics[scale=.79,width=.79\linewidth]{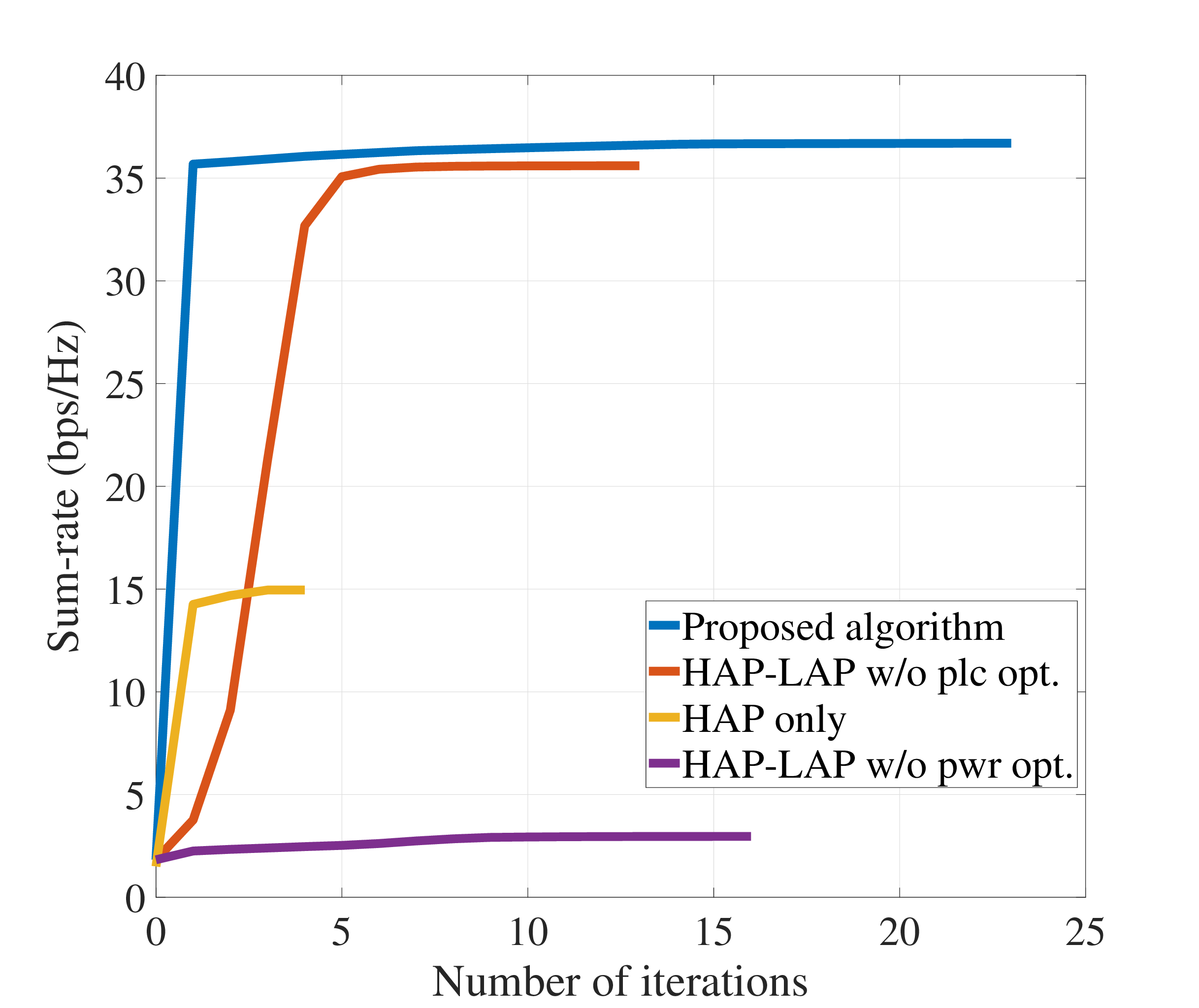}
    \caption{Convergence of Algorithm 1 $(L=4, K=10, C_T=50\textrm{ bps/Hz}$ and $\epsilon = 10^{-3})$.}
        \vspace{-0.6cm}
    \label{Fig_2}
\end{figure}

Fig. \ref{Fig_2} shows the convergence of the proposed algorithm. Compared to other schemes, the proposed algorithm converges fast to the highest sum-rate due to the joint optimization. Also, it is noted that in the case of HAP-LAP w/o pwr opt., the converged sum-rate is incomplete because of the difficulty of interference management, yielding the slow convergence. 

\begin{figure}[t]
  \centering
  \begin{tabular}{c @{\hspace{-10pt}} c }
    \hspace{-8pt}\includegraphics[width=.54\columnwidth, height = 0.54\columnwidth]{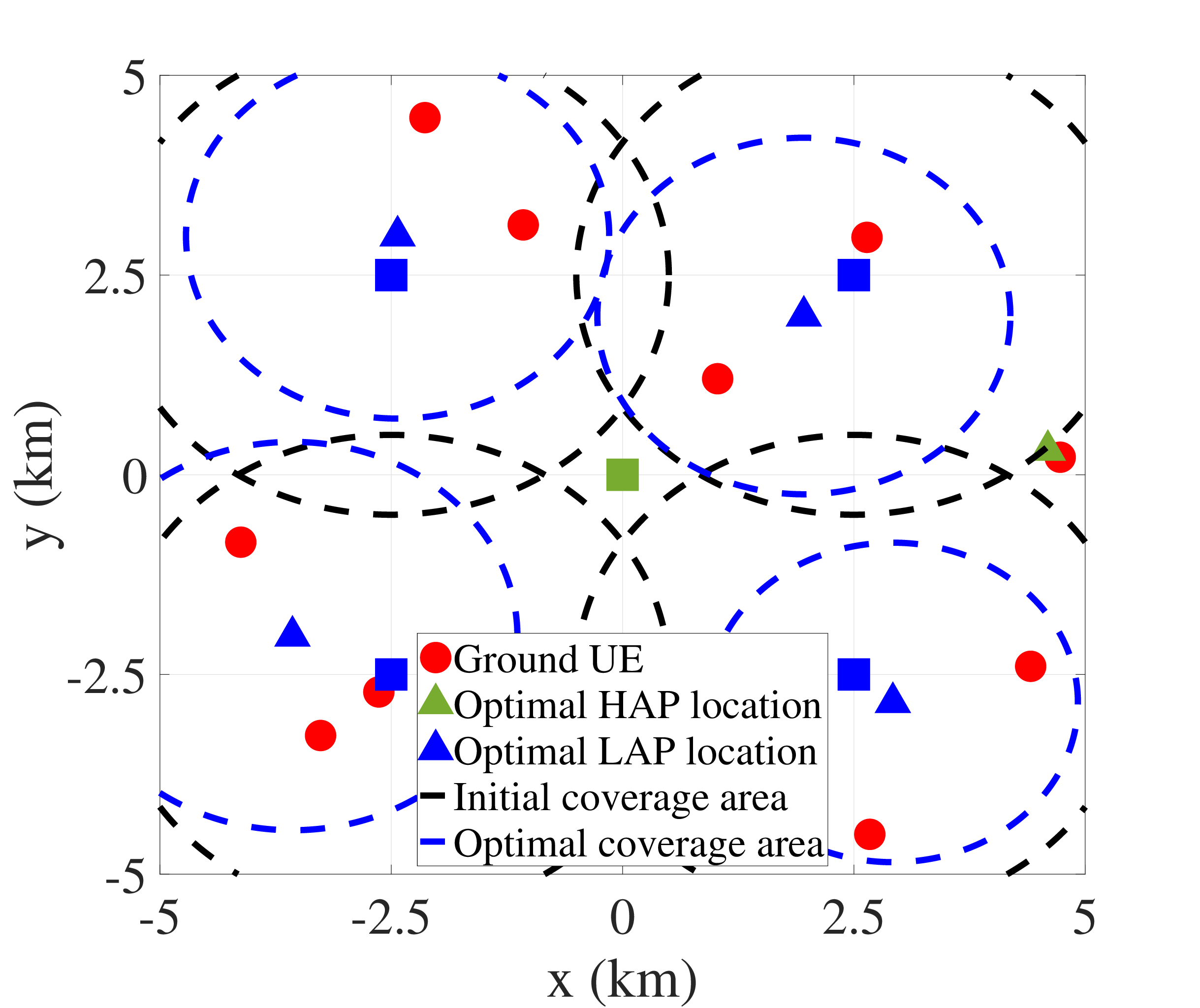} &
      \includegraphics[width=.54\columnwidth, height = 0.54\columnwidth]{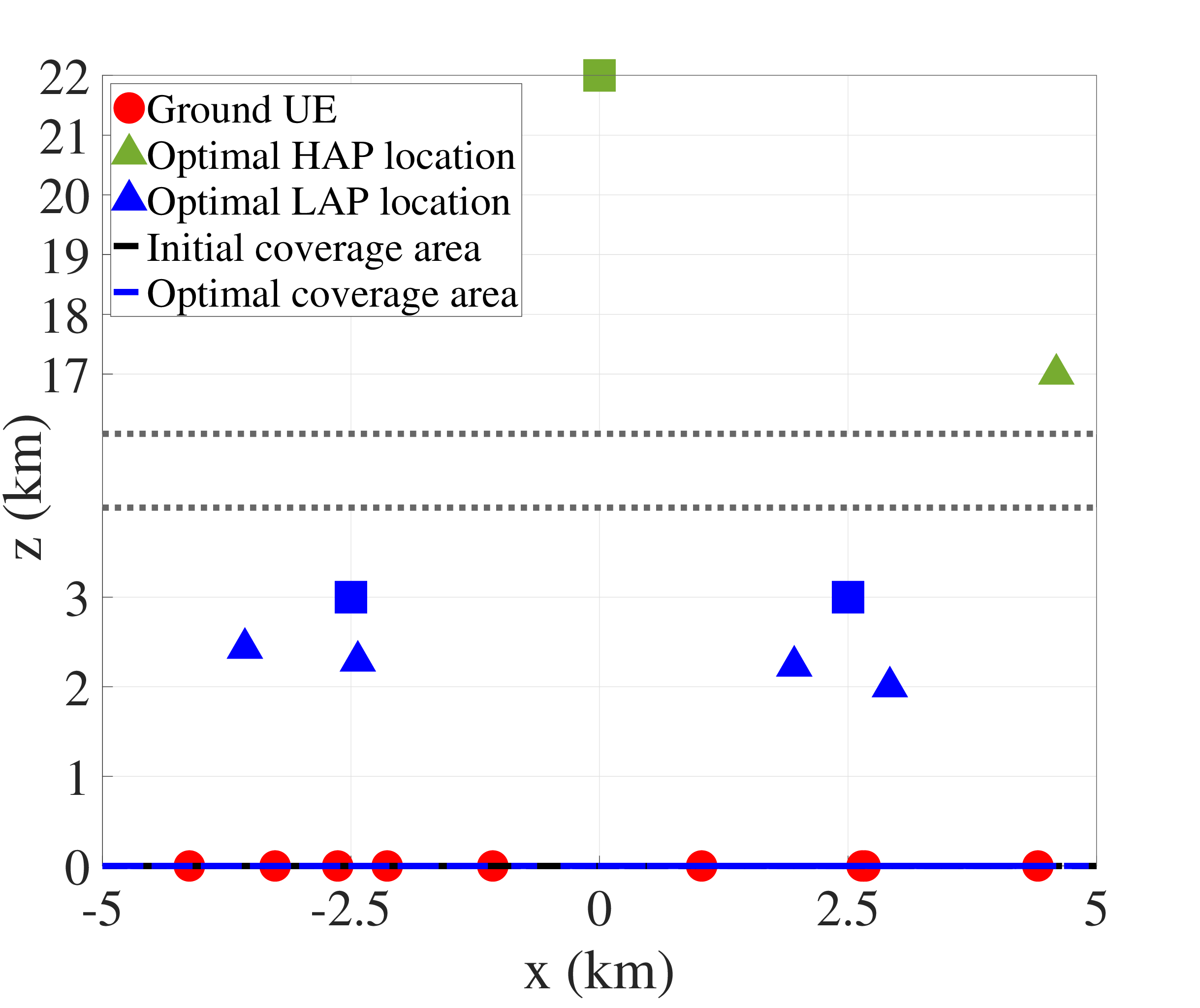} \\
  \end{tabular}
\caption{The optimal UAV locations of xy-plane and xz-plane by the proposed
algorithm $(L=4, K=10, C_T=50\textrm{ bps/Hz}$ and $\epsilon=10^{-3})$.}
        \vspace{-0.5cm}
    \label{Fig_3}
\end{figure}

Fig. \ref{Fig_3} illustrates the optimal UAV placements obtained by proposed Algorithm 1, with initial UAV locations marked as `$\square$' and optimal UAV locations marked as `$\triangle$'. In Fig. \ref{Fig_3}, the optimized altitude of HAP is the minimum allowed value, while the altitude of the LAPs progressively rises. This is because all UEs initially decode the common signal transmitted by the HAP, due to the stronger signal power of HAP with the higher transmit power. In addition, the HAP tends to be closer to the UE outside the LAP coverage areas for satisfying the minimum rate requirements. In the aspect of maximizing the common message rate, the LAPs are located in the way to ensure the equal distance to all connected UEs.

\begin{figure}[t]
  \centering
   \includegraphics[scale=.79,width=.79\linewidth]{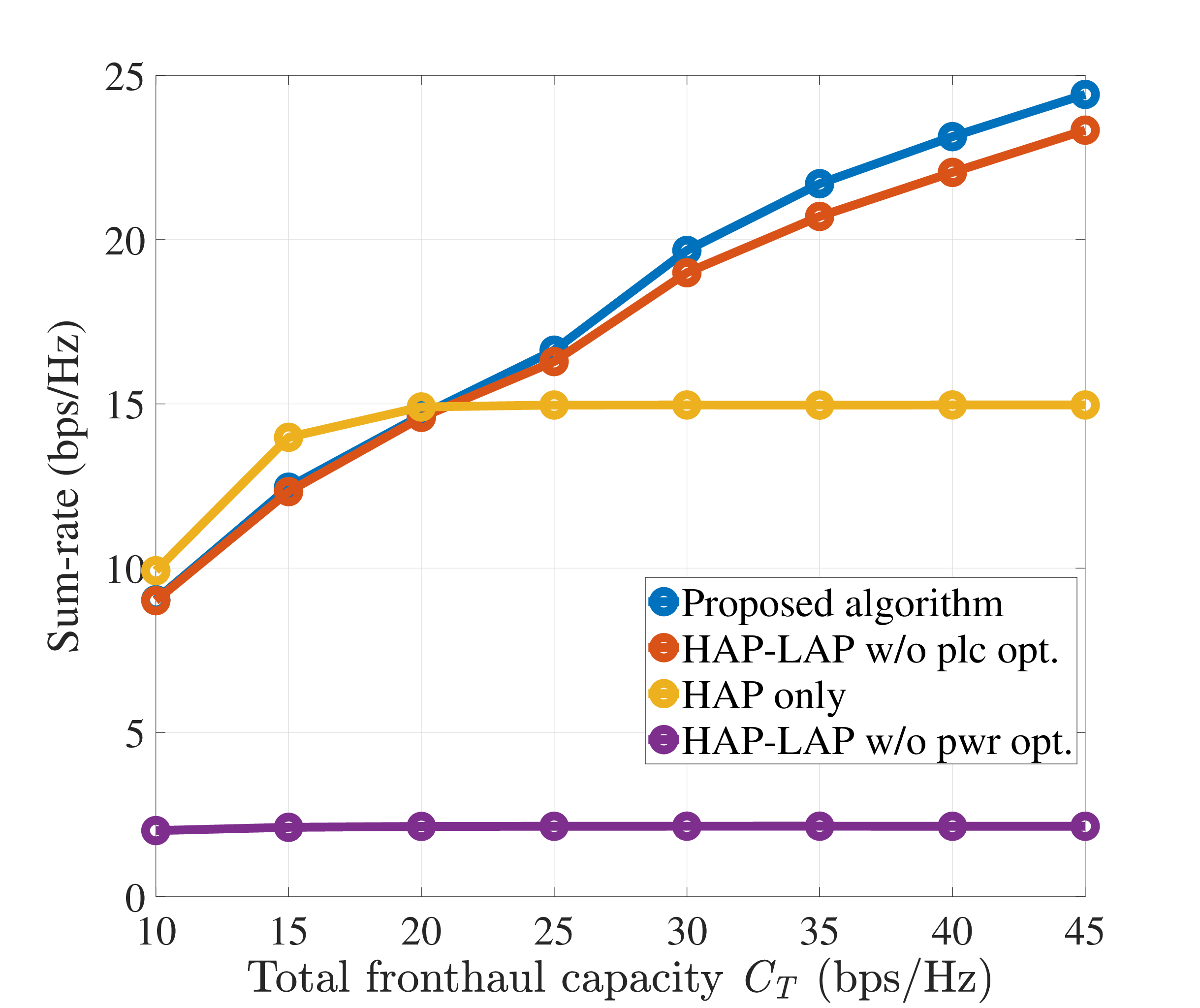}
    \caption{Sum-rate versus the total fronthaul capacity $(L=2, K=6$ and $\epsilon = 2\times 10^{-3})$.}
        \vspace{-0.6cm}
    \label{Fig_4}
\end{figure}

Fig. \ref{Fig_4} plots the sum-rate with respect to the total fronthaul capacity $C_T$. Even though the total fronthaul capacity increases, HAP-LAP w/o pwr opt. achieves only slight improvement in performance, from which the importance of power optimization is pronounced. For the sufficient fronthaul capacity, e.g., $C_T>20$ bps/Hz, the optimal deployment of HAP and LAPs significantly improves the sum-rate performances by effectively utilizing the communication resources. Interestingly, with the insufficient fronthaul, the stand-alone HAP provides the higher sum-rates than that of HAP-LAP networks resulting from the loss on the interference managements. It is concluded that the appropriate deployments of HAP and LAPs along with the optimal resource allocation becomes critical under the limited fronthaul capacity.     

\vspace{-0.15cm}
\section{Concluding remarks}
In this study, we have explored HAP-LAP networks with RSMA under limited fronthaul. Leveraging the C-RAN architecture, we conduct joint optimization of rate-splitting, transmit power, quantization noise variance, and UAV placement. The objective is to maximize the sum-rate of UEs, and through simulations, we have validated the superior performance of our approach compared to benchmark schemes.
As part of our future work, we plan to delve into the UE association and resource allocation of UAVs equipped with multiple antennas.

% if have a single appendix:
%\appendix[Proof of the Zonklar Equations]
% or
%\appendix  % for no appendix heading
% do not use \section anymore after \appendix, only \section*
% is possibly needed

% use appendices with more than one appendix
% then use \section to start each appendix
% you must declare a \section before using any
% \subsection or using \label (\appendices by itself
% starts a section numbered zero.)
%

\vspace{-0.15cm}
\bibliographystyle{IEEEtran}
\bibliography{ref}

%\begin{IEEEbiography}{Michael Shell}
%Biography text here.
%\end{IEEEbiography}

% if you will not have a photo at all:
%\begin{IEEEbiographynophoto}{John Doe}
%Biography text here.
%\end{IEEEbiographynophoto}

% insert where needed to balance the two columns on the last page with
% biographies
%\newpage

%\begin{IEEEbiographynophoto}{Jane Doe}
%Biography text here.
%\end{IEEEbiographynophoto}

% that's all folks
\end{document}